\documentclass[a4paper]{article}
\usepackage{graphicx}
\usepackage{latexsym,amssymb}
\usepackage{amsmath,amscd}

\def\al{\alpha}
\def\ba{\beta} 
\def\ga{\gamma}
\let\da\delta

\def\mbb{\mathbb} 
 
\def\mfr{\mathfrak}
\newtheorem{prop}{Proposition}\def\PRO{\begin{prop}}\def\ORP{\end{prop}}
\newtheorem{coro}{Corollary}\def\COR{\begin{coro}}\def\ROC{\end{coro}}
\newtheorem{theo}{Theorem}\def\TH{\begin{theo}}\def\HT{\end{theo}}
\newtheorem{defi}[prop]{Definition}\def\DE{\begin{defi}}\def\ED{\end{defi}}
\newtheorem{lemme}[prop]{Lemma}\def\LE{\begin{lemme}}\def\EL{\end{lemme}}
\newcommand{\AR}[2][c]{$$\begin{array}[#1]{lllllllllllllll}#2\end{array}$$}
\def\MA#1{\left(\begin{matrix}#1\end{matrix}\right)}
\def\EQ#1{\begin{eqnarray}#1\end{eqnarray}}
\def\mypar#1{\medskip\par\textbf{\emph{#1}}}

\def\ens#1{\{#1\}}

\def\qed{$\Box$}
\def\ei#1{e^{i#1}}
\def\emi#1{e^{{-i}#1}}

\def\G{J}
\def\Rr#1#2{\G_{#1}^{#2}}
\def\ket#1{{|}#1\rangle}
\def\ctR{\mathop{\wedge}\hskip-.4ex} 
\def\ctwo{{\mbb C}^2}
\def\ztwo{{\mbb Z}_2}
\def\ost{\frac1{\sqrt2}}
\def\pit{\frac\pi2}
\def\Cx#1{\cx{#1}{}}
\def\Cz#1{\cz{#1}{}}
\def\cz#1#2{Z_{#1}^{#2}}
\def\cx#1#2{X_{#1}^{#2}}

\def\Ms#1#2{{M}_{#2}^{#1}}
\def\M#1#2{{M}_{#2}^{#1}}
\def\et#1#2{\ctR Z_{#1#2}}

\title{Robust and parsimonious realisations\\ of unitaries in the one-way model}

\author{Vincent Danos\\Universit\'e Paris~7 \& CNRS\\
  {\small\texttt{Vincent.Danos@pps.jussieu.fr}}
  \and
  Elham Kashefi\thanks{This work was partially supported by the PREA, MITACS, ORDCF and CFI projects.}\\ IQC - University of Waterloo\\
  Christ Church College - Oxford\\
  {\small\texttt{ekashefi@iqc.ca}}
  \and
  Prakash Panangaden\\ McGill University\\
  {\small\texttt{Prakash@cs.mcgill.ca}}}

\begin{document}
\maketitle

\begin{abstract}
\noindent

We present a new set of generators for unitary maps over 
$\otimes^n\ctwo$ 
which differs from the traditional rotation-based generating set in
that it uses a single-parameter family of 1-qubit unitaries $\G(\al)$,
together with a single 2-qubit unitary, $\ctR Z$.

Each generator is implementable in the one-way model~\cite{mqqcs} using
only two qubits, and this leads to both parsimonious and robust implementations of general unitaries. As an illustration, we give an implementation of the controlled-$U$ family which uses only 14 qubits, and has a 2-colourable underlying entanglement graph (known to yield robust entangled states).

\end{abstract}

\mypar{Keywords:} Unitary transformations, measurement-based quantum computing.

\section{Introduction}

A parameterised family $\G(\al)$, is shown to generate all unitaries over
$\ctwo$. Adding the unitary operator $\ctR Z$ defined over
$\ctwo\otimes\ctwo$, one obtains a set of generators for all unitary maps
over $\otimes^n\ctwo$, for any $n\in\mbb N$. In other words any
unitary transformation in $\otimes^n\ctwo$ is in the closure of the
set $\ens{\ctR Z, \G(\al); \al\in[0,2\pi]}$ under composition and
tensoring.  This property is commonly referred to as
\emph{universality}. 
One also obtains arbitrarily close approximations using only $\G(0)$, $\G(\frac\pi4)$ and $\ctR Z$.

This special choice of generators, which - as far as we know - was never
considered before, was obtained by looking at the simplest possible
1-qubit measurements patterns in the one-way model~\cite{mqqcs}.  Indeed
both  $\ctR Z$ and $\G(\al)$ may be implemented by one-way measurement
processes, written $\ctR \mfr Z$ and $\mfr \G(\al)$, using only two qubits
each. Therefore, any unitary can be implemented by a combination of $\ctR
\mfr Z$ and $\mfr \G(\al)$. 

In the first section below, we define $\G(\al)$ and $\ctR Z$ and prove that
they are universal.  This first contribution is of general interest.  The
next section turns to implementation matters which are more of specific
quantum computing interest.  We develop a notation for measurement
processes, which we use to describe measurement processes using two qubits
each, and implementing the generators of the first section.  These processes
are proved to be universal in the sense explained above.  Note that the general unitaries implementations are correct by construction, since these are obtained by combining basic processes using composition and tensoring.

As a way of illustrating the low qubit-cost of this representation of
unitaries via measurement processes, we show an implementation of the
controlled-$U$ family using only 14 qubits. It turns out that the
underlying entanglement graph for controlled-$U$ has no odd-length cycle.  
Such specific graphs have been recently shown to yield entangled states which  are robust against decoherence~\cite{DAB03}, and we show that any general unitary can be implemented in the one-way model using one of these.

\section{A universal set for unitaries}
We write $X$ and $Z$ for the Pauli spin matrices:
\AR{
X&:=&\MA{0&1\\1&{0}}&\quad&
Z&:=&\MA{1&0\\0&{-1}}
}
and $H$ and $P(\al)$ for {Hadamard} and {phase} operator:
\AR{
H&:=&\ost\MA{1&1\\1&{-1}}&\quad&
P(\al)&:=&\MA{1&0\\0&\ei\al}
}

\label{s-universal1}
\subsection{A universal set for unitaries on $\ctwo$}
We prove first that the following one-parameter family $\G(\al)$ generates
all unitary operators on $\ctwo$: 
\AR{ 
\G(\al)&:=&\ost\MA{1&\ei\al\\1&-\ei\al}
}

We can see already that the four operators above can be described using only
$\G(\al)$:
\AR{
X&=&\G(\pi)\G(0)\\
Z&=&\G(0)\G(\pi)\\
H&=&\G(0)\\
P(\al)&=&\G(0)\G(\al)
}

We will also use the following equations:
\AR{
\G(0)^2&=&I\\
\G(\al)\G(0)\G(\ba)&=&\G({\al+\ba})\\
\G(\al)\G(\pi)\G(\ba)&=&\ei\al Z\,\G({\ba-\al})
}
The third and fourth one are referred to as the \emph{additivity} and 
\emph{subtractivity} relations. Additivity gives another useful pair of
equations:
\EQ{\label{Xabsorb}
X\G(\al)=\G(\al+\pi)=\G(\al)Z
}

\LE[$\G$--decomposition] 
Any unitary operator $U$ on $\ctwo$ can be written:
\EQ{ U=e^{i\al}\G(0)\G(\ba)\G(\ga)\G(\da)}
for some $\al$, $\ba$, $\ga$ and $\da$ in $\mbb R$.
\EL 
All three Pauli rotations~\cite[p.174]{qcqi} 
are expressible in terms of $\G$: 
\EQ{
\label{xrot}
R_x(\al)&=&\emi{\frac\al2}\G(\al)\G(0)\\
\label{yrot} 
R_y(\al)&=&\emi{\frac\al2}\G(0)\G(\pit)\G(\al)\G(-\pit)\\
\label{zrot}
R_z(\al)&=&\emi{\frac\al2}\G(0)\G(\al) 
}

From the \emph{$Z$--$X$} decomposition, 
we know that every 1-qubit unitary operator $U$ can be written as: 
\AR{ 
U&=&\ei\al R_z(\ba)R_x(\ga)R_z(\da) 
} 
and using equations (\ref{zrot}) and (\ref{xrot}) we get:
\AR{ 
U&=&
\ei\al\emi{\frac{\ba+\ga+\da}2}\G(0)\G(\ba)\G(\ga)\G(\da)
}
\qed

The $Z$--$Y$ decomposition yields the same lemma.

Note that in place of $\G(\al)$, one could take:
\AR{
\ost
\MA{
\emi{\frac\al2}&-\ei{\frac\al2}\\
\emi{\frac\al2}&\ei{\frac\al2}
}
}
which has determinant 1 and leads to similar equations, but is slightly less convenient for computations. One could also have
both $P(\al)$ and $H$ as generators. However the smallest implementation of $P(\al)$ in the one-way model costs 3 qubits, whereas, as we will see in the next section, $\G(\al)$ only costs 2 qubits. 

\subsection{A universal set for unitaries on $\otimes^n\ctwo$}
Now that we know that $\G(\al)$ generates all 1-qubit unitary operators,
we turn to the question of generating unitaries of any finite dimension. 

Given $U$ a unitary over $\ctwo$, one defines a new unitary on $\ctwo\otimes\ctwo$, written $\ctR U$ and read \emph{controlled-$U$}, 
as follows:
\AR{
\ctR U\ket{0}\ket{\psi}&:=&\ket{0}\ket{\psi}\\
\ctR U\ket{1}\ket{\psi}&:=&\ket{1}U(\ket{\psi})
}
We need a second generator to complete our set:
\AR{ 
\ctR{Z}&:=&\MA{1&0&0&0\\0&1&0&0\\0&0&1&0\\0&0&0&-1}
}
The next lemma gives the decomposition of $\ctR U$, in terms of $\G(\al)$
and $\ctR Z$. Subscripts to operators indicate the qubit to which they
apply, and we sometimes abbreviate $\G_i(\al)$ as $\Rr i\al$.

\LE\label{controlledU}
For any unitary operator $U$ on $\ctwo$ with $\G$-decomposition:
\AR{\ei\al\G(0)\G(\ba)\G(\ga)\G(\da)}
one has:
\AR{ 
\ctR U_{12}
&=& 
\Rr{1}{0}\Rr{1}{\al'}
\Rr{2}{0}\Rr{2}{\ba+\pi}
\Rr2{-\frac\ga2}\Rr2{-\pit}
\Rr20\ctR Z_{12}
\Rr2{\pit}\Rr2{\frac\ga2}
\Rr2{\frac{-\pi-\da-\ba}2} 
\Rr20
\ctR Z_{12}
\Rr2{\frac{-\ba+\da-\pi}2} 
}
with $\al'=\al+\frac{\ba+\ga+\da}2$.
\EL

Define auxiliary unitary operators:
\AR{
A&=&\G(0)\G(\ba+\pi)\G(-\frac\ga2)\G(-\pit)\\
B&=&\G(0)\G(\pit)\G(\frac\ga2)\G(\frac{-\pi-\da-\ba}2)\\
C&=&\G(0)\G(\frac{-\ba+\da-\pi}2) 
} 
Using the additivity relation, it is easy to verify that $ABC=I$. 

On the other hand, using both the subtractivity relation and
equations~(\ref{Xabsorb}), we get:

\AR{
AXBXC&=&
\G(0)\G(\ba+\pi)\G(-\frac\ga2)\G(-\pit)
\G(\pi)\G(\pit)\G(\frac\ga2)
\G(\frac{-\pi-\da-\ba}2)\G(\pi)\G(\frac{-\ba+\da-\pi}2)
\\
&=&
\G(0)\G(\ba+\pi)\G(-\frac\ga2)
\emi{\pit}Z
\G(\pi)
\G(\frac\ga2)
\ei{\frac{-\pi-\da-\ba}2}Z
\G(\da)\\
&=&
-\ei{\frac{-\da-\ba}2}
\G(0)\G(\ba+\pi)
\G(-\frac\ga2+\pi)
\G(\pi)
\G(\frac\ga2+\pi)
\G(\da)\\
&=&
-\ei{\frac{-\da-\ba}2}
\G(0)\G(\ba+\pi)
\ei{(-\frac\ga2+\pi)}Z
\G(\ga)
\G(\da)\\
&=&
\emi{\frac{\da+\ba+\ga}2}
\G(0)\G(\ba)
\G(\ga)
\G(\da)\\
}
Therefore one also has $\ei{\frac{2\al+\ba+\ga+\da}2} AXBXC=U$.

Combining our two equations in $A$, $B$, $C$, we obtain the following
decomposition of $\ctR U$ 
(recall that subscripts indicate the qubits on which the operators act):  
\AR{
\ctR U_{12}&=& P_1(\al')A_{2}\ctR X_{12} B_{2}\ctR X_{12} C_{2}
}
with $\al'=\al+\frac{\ba+\ga+\da}2$;
a decomposition which we can rewrite using our generating set:
\AR{
P(\al)_{1}&=&\Rr10\Rr1{\al}\\
\ctR X_{12}&=&H_{2}\ctR Z_{12}H_{2}&=& \Rr20\ctR Z_{12}\Rr20} 
to establish the lemma.
\qed

As we will see, this expression leads to an implementation for the $\ctR U$ operator using only 14 qubits. Using the $Z$--$X$ decomposition, one finds another decomposition costing 15 qubits, while
the $X$--$Y$ decomposition costs 16 qubits. We haven't found any expression leading to a cheaper implementation.

Having all unitaries $U$ over $\ctwo$ and all unitaries of the form $\ctR U$
over $\ctwo\otimes\ctwo$ we can conclude to universality~\cite[pp.~191--194]{qcqi}:
\TH[Universality] 
The set $\ens{\G(\al),\ctR Z}$ generates all unitaries. 
\HT 

The following unitaries $H=\G(0)$,
$P(\frac\pi4)=\G(0)\G(\frac\pi4)$, and 
$\ctR X=\G(0)\ctR Z\G(0)$, are known to be \emph{approximately universal},
in the sense that any unitary 
can be approximated within any precision by combining
these~\cite[p. 197]{qcqi}. Therefore the set $\G(0)$, $\G(\frac\pi4)$ and $\ctR Z$ is also approximately universal.

\section{One-way implementations}
So far we have obtained a generating set for unitaries which compares well
with the usual one based on $\ctR X$ and general rotations
$R(\al,\ba,\ga)$, in that our 1-qubit family of unitaries has only one
parameter.  As we have noted earlier, another advantage of our generating
set is that 
$\G(\al)$ and $\ctR Z$ have realisations in the one-way model using fewer
qubits. This is the topic of this section. 

Measurement based models~\cite{L03,AL04,CLN04} have been recently brought
to the fore, notably the \emph{one-way} model~\cite{mqqcs}. One reason why
this universal model is  
particularly interesting is that it might lead to easier
implementation~\cite{BR04,CMJ04,ND04}. 

\subsection{The one-way model}
We begin with a quick recapitulation of the one-way model. Computations
involve combinations of 2-qubit entanglement operators $\et ij$, 1-qubit
measurements
$\M\al i$, and 1-qubit Pauli corrections $\Cx i$, $\Cz
i$, where $i$, $j$ represent the qubits on which each of these operations
apply, and $\al$ is a parameter in $[0,2\pi]$.  Such combinations, together
with two distinguished set of qubits (possibly overlapping) corresponding to
inputs and outputs, will be
called \emph{measurement patterns}, or simply patterns.
One can associate an \emph{entanglement graph} to any pattern,
where the vertices are the pattern qubits, 
and the (undirected) edges are given by the $\et ij$ operators~\cite{graphstates}.

Importantly, in a pattern, corrections are allowed to depend on previous
measurement outcomes. There is also a parallel notion of dependent
measurements, but we will not use it here.

To be more specific, $\M\al i$ is an $xy$-measurement, defined by a pair of
complement orthogonal projections, applied at qubit $i$, on the following
vectors:  
\EQ
{
\ket{+_{\al}}=\ost(\ket0+\ei{\al}\ket1),\,
\ket{-_{\al}}=\ost(\ket0-\ei{\al}\ket1)
}
Conventionally, we take measurement outcomes to range in $\ztwo=\ens{0,1}$,
$0$ corresponding 
to a collapse to $\ket{+_{\al}}$, and $1$ to a collapse to $\ket{-_{\al}}$.

Since qubits are measured at most once, we may
represent unambiguously the outcome of the measurement done at qubit $i$ by
$s_i$. 
Dependent correction will be written $\cx is$ and $\cz is$ with 
$s=\sum_{i\in I} s_i$. Their meaning is that $\cx i0=\cz i0=I$ (no
correction is applied), while 
$\cx i1=\Cx i$ and $\cz i1=\Cz i$.

Now that we have a notation, we may describe the two patterns implementing
our generators: 
\EQ{
\mfr \G(\al)&:=&\cx 2{s_1}\M{{-\al}}1\et 12\\
\ctR{\mfr Z}&:=&\et 12 
}
In the first pattern $1$ is the only input and $2$ is the only output,
while in the second both $1$ and $2$ are inputs and outputs (note that we are allowing patterns to have overlapping inputs and outputs).

These patterns are indeed among the simplest possible.  Remarkably, there
is only one single dependency overall, which occurs in the correction phase
of $\mfr \G(\al)$.  No set of patterns without any measurement could be a
generating set, since those can only implement unitaries in the Clifford
group.  Dependencies are also likely to be needed for universality, but for
the moment we do not have a proof for this.

It is easy to verify that $\mfr \G(\al)$ and $\ctR{\mfr Z}$ respectively
implement $\G(\al)$ and $\ctR{Z}$.  Combining these two patterns, by
composition and tensoring, will therefore generate patterns realising all
unitaries over $\otimes^n\ctwo$.

\subsection{Controlled-$U$ implementation}
\begin{figure}[h]
\begin{center}
\includegraphics[scale=0.6]{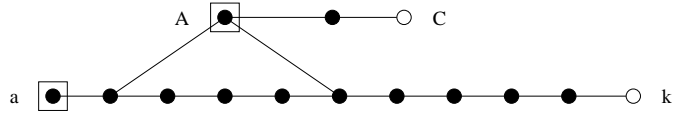}
\caption{
\footnotesize{Graph state for the $\ctR U$ pattern}: inputs $A$, $a$ are represented as boxes, outputs $C$, $k$ as empty circles, measured qubits as solid circles.}
\label{controlled-U}
\end{center}
\end{figure}

To conclude, we describe the entanglement graph needed for $\ctR U$,
according to the decomposition given in lemma~\ref{controlledU}. As we
see in Figure~\ref{controlled-U} this graph has a single cycle of length~6. (Others implementations, based on other decompositions, could use a different graph.)

Recall that the controlled-$U$ decomposition is:
\AR{ 
\ctR U_{12}
&=& 
\Rr{1}{0}\Rr{1}{\al'}
\Rr{2}{0}\Rr{2}{\ba+\pi}
\Rr2{-\frac\ga2}\Rr2{-\pit}
\Rr20\ctR Z_{12}
\Rr2{\pit}\Rr2{\frac\ga2}
\Rr2{\frac{-\pi-\da-\ba}2} 
\Rr20
\ctR Z_{12}
\Rr2{\frac{-\ba+\da-\pi}2} 
}
with $\al'=\al+\frac{\ba+\ga+\da}2$.

Replacing each of the generators in the above expression
by the corresponding module, we get:
\AR{
\cx{C}{s_B}\Ms{0}{B}{}{}\et{B}{C}
\cx{B}{s_A}\Ms{-\al'}{A}{}{}\et{A}{B}
\cx{k}{s_j}\Ms{0}{j}{}{}\et{j}{k}
\cx{j}{s_i}\Ms{-\ba-\pi}{i}{}{}\et{i}{j}
\\
\cx{i}{s_h}\Ms{\frac\ga2}{h}{}{}\et{h}{i}
\cx{h}{s_g}\Ms{\pit}{g}{}{}\et{g}{h}
\cx{g}{s_f}\Ms{0}{f}{}{}\et{f}{g}
\et{A}f
\cx{f}{s_e}\Ms{-\pit}{e}{}{}\et{e}{f}
\\
\cx{e}{s_d}\Ms{-\frac\ga2}{d}{}{}\et{d}{e}
\cx{d}{s_c}\Ms{\frac{\pi+\da+\ba}2}{c}{}{}\et{c}{d}
\cx{c}{s_b}\Ms{0}{b}{}{}\et{b}{c}
\et{A}b
\cx{b}{s_a}\Ms{\frac{\ba-\da+\pi}2}{a}{}{}\et{a}{b}
}
with inputs $\ens{A,a}$ and outputs $\ens{C,k}$.
We have reserved uppercase (lowercase) letters for qubits
used in implementing $\G$s on control qubit $1$ (target qubit $2$),
and used a total number of 14 qubits as expected.

\subsection{Robust implementations}
\begin{figure}[h]
\begin{center}
\includegraphics[scale=0.6]{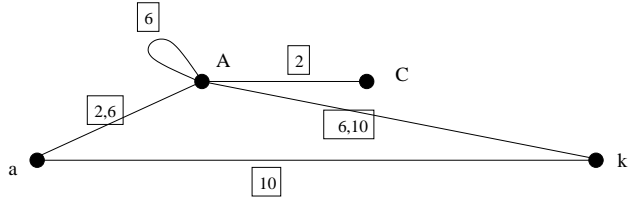}
\caption{
\footnotesize{Extreme paths in $\ctR U$ pattern}: numbers represent the length of paths; solid circles represent the pattern inputs and outputs.}
\label{paths}
\end{center}
\end{figure}

The pattern for $\ctR U$ has a further interesting property,
namely that all possible paths linking boundary vertices (inputs
and outputs) are of even length (2, 6, 10 as it happens) as we can see
on Figure~\ref{paths}.

Say a path is \emph{extreme} in a graph with inputs and outputs,
if it goes from the boundary to itself; say a graph
with inputs and outputs is \emph{even} if all its
extreme paths are of even length; say a pattern
is even if its entanglement graph is. 
We may then rephrase the last observation by saying that
the pattern for $\ctR U$ is even.

Patterns with empty command sequence, among which one finds those implementing permutations over $\otimes^n\ctwo$, are even. Indeed, all their paths, therefore all their extreme paths are of length zero, and zero is even.

\LE 
Even patterns are closed under tensorisation and composition.
\EL
The first point is clear, because the graph associated to a tensor of two patterns is the juxtaposition of the components graphs, and therefore it has a path space which is the disjoint sum of the path spaces of its components.

For the second point, we note that any extreme path in a composite pattern is a product of extreme paths of the components, and has therefore even length,
because the sum of evens is even.
\qed

\PRO
Any unitary can be realised by a pattern with 
a 2-colourable underlying graph.
\ORP
Any unitary can be realised by a pattern obtained from the $J$-decomposition pattern (using 5 qubits), the $\ctR U$
pattern, and the permutation patterns, combined by tensor and composition. 
Any cycle in such a pattern is (1) either a cycle internal to some basic pattern (which rules out $J$-decomposition and permutation patterns which have a linear entanglement graph), so living inside a $\ctR U$ pattern, therefore of length $6$, therefore even, as we have seen,
or (2) a product of extreme paths, therefore even, because all basic patterns are even, and by the lemma above, so is any combination of them.
\qed

As said in the introduction, 2-colourable entanglement graphs are interesting 
since their associated graph states are robust against decoherence~\cite{DAB03}.

\section{Conclusion}
We have shown a simple set of generators for the group of unitary maps over 
$\otimes^n\ctwo$ which has not been considered before as far as we know.
This set might be of general interest for the understanding and manipulation of 
the unitary group. 

However, our primary interest was to use it as a way to
understand the measurement-based quantum computing  model known as the
one-way model~\cite{mqqcs}. Indeed, this set yields parsimonious 
implementations of unitaries in the one-way model. 
Moreover, we were able to show that the entanglement graphs underlying these implementations can always be chosen to be 2-colourable. Such graphs lead to entangled states belonging to the family of graph states for which physical implementations are well underway~\cite{BR04,CMJ04}.  
Purification protocols exist for the particular subclass of 2-colourable graph states, which make them physically implementable in a way that is robust against decoherence~\cite{DAB03}.

\end{document}